# Sub-nanosecond in-plane magnetization switching induced by field-like spin-orbit torques from ferromagnets

Hanying Zhang[1, 2], Ziqian Cui[1, 2], Baiqing Jiang[1, 2], Yuan Wang[1, 2], and C. Bi[1, 2]*

[1]Laboratory of Microelectronics Devices and Integrated Technology, Institute of Microelectronics, Chinese Academy of Sciences, Beijing 100029, China

[2]University of Chinese Academy of Sciences, Beijing, 100049, China

*clab@ime.ac.cn

**Abstract:**

Spin-orbit torques (SOTs) generated in SOT-material/ferromagnet structures are classified as damping-like (DL) and field-like (FL) torques for current-driven magnetization switching. It is well known that both DL- and FL-SOTs originate from the SOT-material and DL-SOT dominates the current-driven switching process while FL-SOT contributes limitedly, resulting in an incubation time (several nanoseconds) during collinear magnetization switching with the spin polarization because of the DL attributes. Here we report a FL-SOT originated from the ferromagnet, different from the origin of DL-SOT, and demonstrate that it dominates the collinear magnetization switching. We show that the FL-SOT and resultant collinear switching can be modulated, one order of magnitude and sign reversal, by controlling the ferromagnet. Because of no incubation time and higher charge-to-spin efficiencies in the FL switching, we further show that the switching time can be down to 200 ps with one order lower critical switching current density compared to DL switching. These results indicate that the FL switching may provide a practical solution for magnetic memory in speed-priority cache applications.

Spin-orbit torques (SOTs) generated at SOT-material/ferromagnet interfaces have been proposed to manipulate the ferromagnet like spin-transfer torque (STT), which can be used to develop low-power spin-based logic and memory devices [1–6]. SOT arises from the spin polarization ($\sigma$) at the SOT-material/ferromagnet interfaces due to the spin Hall (SHE) or Rashba effects, as schematically shown in Fig. 1(a). It can be classified into damping-like (DL) and field-like (FL) torques depending on how to interact with the ferromagnet [4,7,8]. The DL-torque, $\tau_{DL} = C_{DL}J\mathbf{m} \times (\sigma \times \mathbf{m})$, competes with the damping term of Landau-Liftshitz-Gilbert (LLG) equation describing magnetization dynamics during magnetization manipulation, while the FL-torque, $\tau_{FL} = C_{FL}J\mathbf{m} \times \sigma$, acting like a magnetic field. Here, $\mathbf{m}$ is unit magnetization, $J$ is current density, and $C_{FL}$, $C_{DL}$ are two coefficients. Many SOT-materials including heavy-metals [1–6], magnetic metals [9–16], and topological insulators [17–21], have been investigated, but mostly, focused on the $\tau_{DL}$, since it dominates the current-driven magnetization switching both perpendicular to and collinear with $\sigma$ while the $\tau_{FL}$ contributes limitedly [1,2,24–27,3–6,9,15,22,23]. Moreover, it is well known that both $\tau_{DL}$ and $\tau_{FL}$ result from $\sigma$ with the same origin determined by SOT-materials [1,3–5,9,15].

In principle, both $\tau_{DL}$ and $\tau_{FL}$ alone can induce collinear magnetization switching as simulated in Figs. 1(b) and 1(c), where the $\tau_{DL}$-induced magnetization switching requires an incubation time to accumulate nonlinear spin torques in the early switching stage like STT [22], while the $\tau_{FL}$-switching promises a much faster switching process because of no incubation time. However, there are no experiments demonstrating the $\tau_{FL}$-induced magnetization switching so far, and the current-driven collinear switching is attributed to the $\tau_{DL}$-induced switching that was used to determine SHE sign or evaluate the charge-to-spin conversion efficiency for different SOT-materials [1,3–6,9,21]. In this letter, we first show that the $\tau_{FL}$ strongly depends on adjacent ferromagnets, indicating a different origin from $\tau_{DL}$, and then demonstrate that it dominates the current-driven collinear switching. We show that $\tau_{FL}$ can be modulated independently, both the magnitude and sign, while keeping $\tau_{DL}$ sign the same, and thus, the $\tau_{FL}$-induced switching can be clearly determined according to switching directions that are always consistent with $\tau_{FL}$ sign. The high charge-to-spin efficiency and ultrafast switching process down to 200 ps due to the absence of incubation time have also been demonstrated.

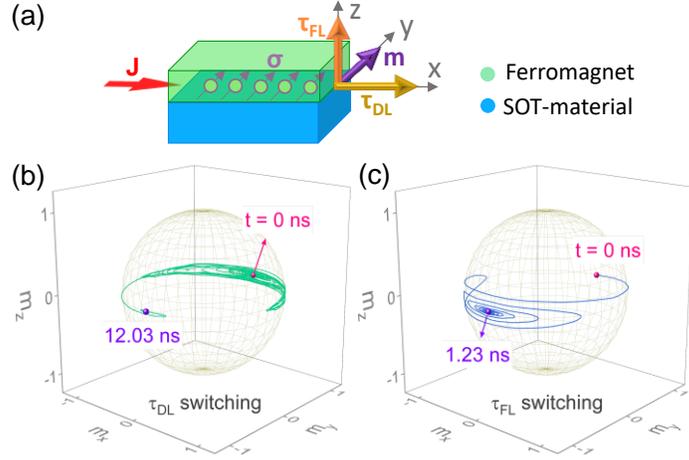

FIG. 1. (a) Schematic of $\tau_{FL}$ and $\tau_{DL}$ generation at SOT-material/ferromagnet interfaces. The three-dimensional magnetization trajectories during $\tau_{DL}$- (b) and $\tau_{FL}$-driven (c) collinear in-plane switching simulated by using the same $\tau_{FL}$ and $\tau_{DL}$ values (Supplementary Materials).

SOT-material/ferromagnet structures consisting of Pt2.5/Ni $t_{Ni}$/CoFeB2.5/MgO (nm) bilayers were chosen to investigate $\tau_{FL}$ and following collinear in-plane switching, where a Ni layer with 0 nm ≤ $t_{Ni}$ ≤ 2 nm was inserted between the Pt and CoFeB layers. The inserted Ni and CoFeB layers can be viewed as a single ferromagnet during SOT characterization and magnetization manipulation because of direct exchange coupling. Therefore, any change of $\tau_{FL}$ with $t_{Ni}$ must arise from ferromagnets, rather than Pt or Pt-modulated interface. CoFeB was adopted because it shows large unidirectional magnetoresistance (USMR) (Supplementary Materials) for detecting in-plane switching [28].

Two independent measurements, a field-compensated approach [29,30] and spin-torque ferromagnetic resonance (ST-FMR) [31], were performed to qualitatively characterize $\tau_{FL}$ (Fig. 2). A Hall bar (2.5 × 50 µm²) was used for the field-compensated measurements (Fig. (2b)) [29,30]. The current (*I*) and external magnetic field ($H_x$) were applied in the *x*-direction, and a small in-plane compensated field ($h_y$) was applied along the *y*-direction. The $\tau_{FL}$ can be viewed as an effective in-plane field ($H_{FL}$) orthogonal to *I*. Combining with the Oersted field ($h_O$), $H_{FL}$ tilts the magnetization away from the *x*-axis with a small angle $\varphi \approx \Delta m_y$, where $\Delta m_y$ is the change of **m** along *y*-direction and $\Delta m_y = 0$ if $H_{FL} + h_O + h_y = 0$. $\Delta m_y$ can be detected by planar Hall resistance ($R_{xy}$), $\Delta R_{xy} = R_{xy}(+I) - R_{xy}(-I) \propto \Delta m_y$. Figure 2(a) shows $\Delta R_{xy}$ (*I* = ±1 mA) as a function of $H_x$ under various $h_y$ for Pt/CoFeB/MgO. When $|H_x| > 1$ kOe, $\Delta R_{xy}$ becomes flat, which is attributed to the anomalous Nernst effect (ANE), $\propto$ **m** × ΔT, with a vertical temperature gradient ΔT. The peaks around zero field are due to $\Delta m_y$ that were modulated by $h_y$ dramatically. When $h_y$ = -1.2 Oe, the peaks disappear and only ANE signal exists, indicating $H_{FL} + h_O +$ $h_y = 0$. Therefore, the estimated $H_{FL}$ is 1.9 Oe along +y direction by using $h_O = -0.7$ Oe (Supplementary Materials). When $h_y = -1.8$ Oe, the peaks reverse, confirming that they do arise from $H_{FL} + h_O + h_y$.

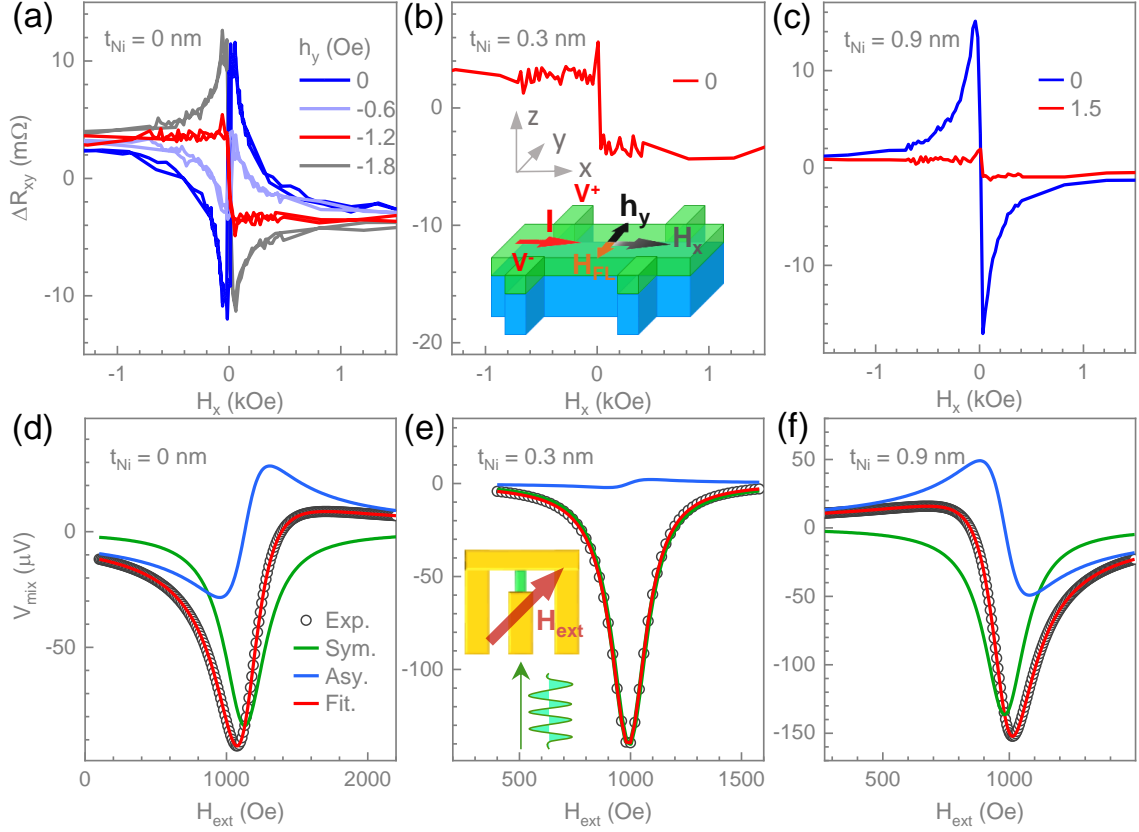

FIG. 2. (a) $\Delta R_{xy}$ as a function of $H_x$ under $h_y$ modulations for Pt/CoFeB/MgO. (b) No peaks observed for Pt/Ni0.3nm/CoFeB/MgO. (c) $\Delta R_{xy}$ versus $H_x$ for Pt/Ni0.9nm/CoFeB/MgO. ST-FMR spectra and corresponding fitting for Pt/CoFeB/MgO (d), Pt/Ni0.3nm/CoFeB/MgO (e), and Pt/Ni0.9nm/CoFeB/MgO (f). Insets: schematic of experimental setups.

Similar $h_y$-modulated $\Delta R_{xy}$ curves were also observed when $t_{Ni} > 0$ nm, and importantly, the fully compensated $h_y$ changes sign with increasing $t_{Ni}$. Figure 2(c) shows $\Delta R_{xy}$ curves for Pt/Ni0.9nm/CoFeB/MgO. Compared to Fig. 2(a), the peaks reverse even no $h_y$ applied, indicating $H_{FL} + h_O$ changes sign. As expected, a positive $h_y$ eliminates the peaks, indicating $H_{FL} + h_O = -1.5$ Oe with 0.9 nm Ni insertion. Because $h_O = -0.7$ Oe, $H_{FL}$ must change sign and equal -0.8 Oe to satisfy $H_{FL} + h_O = -1.5$ Oe ($|h_O| < 0.7$ Oe and $H_{FL} < -0.8$ Oe ($|H_{FL}| > 0.8$ Oe) if considering current shunt of Ni). This is different from the case that $H_{FL} + h_O$ changes sign induced by $H_{FL} \approx 0$ Oe with a thick ferromagnet [29,32,33]. Note that the ANE signal of Fig. 2(c) ($|H_x| > 1$ kOe) keeps the same sign as Fig. 2(a) since both **m** and $\Delta T$ do not change sign. No peaks shown in the $\Delta R_{xy}$ curves of Pt/Ni0.3nm/CoFeB/MgO (Fig. 2(b)) indicate that H$_{FL}$ and $h_O$ cancel each other, suggesting gradually modified H$_{FL}$ with Ni insertions. These results clearly demonstrate that τ$_{FL}$ strongly depends on ferromagnets and reverses sign with Ni insertions. In SHE mechanisms [34,35], both τ$_{FL}$ and τ$_{DL}$ only originate from SOT-materials, not relating to adjacent ferromagnets; in a pure Rashba scheme [7,27,28], the direction of Rashba field is determined by a net potential gradient ($\nabla V$) due to inversion asymmetry at the interface [36,37] which is not expected to change sign with a few inserted Ni monolayers. Moreover, according to these theories, τ$_{FL}$ is generated through **σ** whether from a SHE [34] or Rashba origin [37,38], which conflicts below USMR results that **σ** does not change sign with Ni insertions, and thus τ$_{FL}$ sign should not change. Therefore, τ$_{FL}$ must arise from ferromagnets, excluding the SHE or pure Rashba mechanisms.

ST-FMR results characterizing τ$_{FL}$ and τ$_{DL}$ simultaneously are presented in Figs. 2(d-f), which can be decomposed into a symmetric and an antisymmetric peak, corresponding to τ$_{DL}$ and τ$_{FL}$, respectively. Figures 2(d-f) show that the antisymmetric peak changes sign when t$_{Ni}$ = 0.9 nm, indicating sign reversal of τ$_{FL}$ (also reported in Pt/Ni/Py and Ta/Ni/Py structures recently [39]) . For Pt/Ni0.3nm/CoFeB/MgO, the antisymmetric peak almost disappears (Fig. 2(e)). These results are fully consistent with τ$_{FL}$ determined in Figs. 2(a-c). Notably, the symmetric peak and thus τ$_{DL}$ shows the same sign regardless of t$_{Ni}$, indicating a SHE origin from SOT-materials. The fact that only τ$_{FL}$ sign reverses provides empirical evidences for distinguishing the following τ$_{FL}$-driven switching where the switching direction is always determined by τ$_{FL}$ signs. Otherwise, ST-FMR results also exclude the possible orbital Hall effects (OHE) as the origin of τ$_{FL}$ and τ$_{DL}$, since both are expected to change sign simultaneously [40]. In Supplementary Materials, we propose that τ$_{FL}$ arises from a ferromagnet-originated Rashba effect due to asymmetric spin-dependent relaxation time, where the single electron Hamiltonian, $\hat{H} = \frac{\hat{\mathbf{p}}^2}{2m_e^*} - J_{ex}\boldsymbol{\sigma} \cdot \mathbf{m}\Theta(z) + \hat{V}_s(z)$ with the scattering potential $\hat{V}_s(z) = (\alpha_R/\hbar)\delta(z)\boldsymbol{\sigma} \cdot (\hat{\mathbf{p}} \times \mathbf{z})$, is characterized by interfacial Rashba spin-orbit coupling, $\alpha_R$. The spin current is evaluated by calculating the reflection and transmission amplitudes of scattering waves and the resultant τ$_{FL}$ depends on asymmetric spin-dependent relaxation time in adjacent ferromagnets (Supplementary Materials).

Figure 3 demonstrates the τ$_{FL}$-induced magnetization switching detected by using USMR and magneto-optical Kerr effects (MOKE). Figure 3(a) shows the USMR signal, $\Delta R_{xx} = R_{xx}(I) - R_{xx}(-I)$, as a function of H$_y$ for Pt/CoFeB/MgO and Pt/Ni0.5nm/CoFeB/MgO. The peaks around zero field reverse in the two samples due to electron-magnon scattering [41] (Supplementary Materials). $\Delta R_{xx}$ saturating at larger fields arises from spin-dependent electron transport [41,42], where the same sign of saturated $\Delta R_{xx}$ confirms that **σ** does not change direction in the two samples. As mentioned above, if both τ$_{FL}$ and τ$_{DL}$ originate from **σ**, their signs should keep the same, which contradicts with the above experimental observations that $\tau_{FL}$ sign reverses with increasing $t_{Ni}$. Therefore, USMR results also support that $\tau_{DL}$ has a Pt-related SHE origin while $\tau_{FL}$ has a ferromagnet-dominated origin.

The inset of Fig. 3(a) shows that $\Delta R_{xx}$ can distinguish the two stable magnetization states clearly at zero field. Note that $\Delta R_{xx}$ sign reverses in the two samples at zero field, which is very important for determining current-induced switching directions. Figure 3(b) presents time-dependent $H_y$ and measured $\Delta R_{xx}$, where the magnetization was aligned to a uniform state by ±1 kOe field. $\Delta R_{xx}$ jumps sharply after removal of $H_y$ (from ±1 kOe to 0 Oe) because of different USMR mechanisms at high (±1 kOe) and low fields [33] even magnetization points the same direction, as explained above. For Pt/CoFeB/MgO, $\Delta R_{xx}$ aligned by –1 kOe field (-M states) stabilizes at a positive value after removing $H_y$, while for Pt/Ni0.5nm/CoFeB/MgO, it stabilizes at a negative value. For +M states, $\Delta R_{xx}$ stabilizes at negative and positive values for Pt/CoFeB/MgO and Pt/Ni0.5nm/CoFeB/MgO, respectively. These results also confirm opposite $\Delta R_{xx}$ signs at zero field in the two samples.

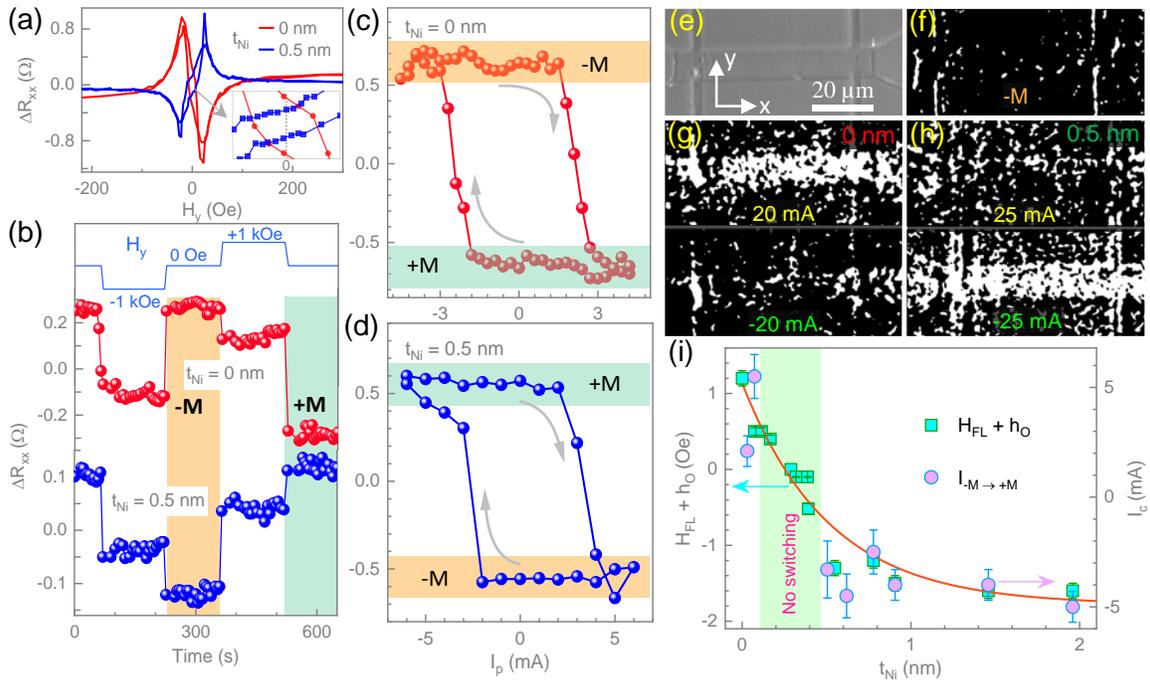

FIG. 3. (a) USMR results for Pt/CoFeB/MgO and Pt/Ni0.5nm/CoFeB/MgO (I: ±1 mA). Inset: enlarged central part. (b) Time dependence of $\Delta R_{xx}$ aligned by $H_y$ for Pt/CoFeB/MgO (middle) and Pt/Ni0.5 nm/CoFeB/MgO (bottom). (c, d) $\Delta R_{xx}$ versus $I_p$ for Pt/CoFeB/MgO (c) and Pt/Ni0.5 nm/CoFeB/MgO (d). (e) Optical images of Hall bars. (f) MOKE images of –M states aligned by -200 Oe field. (g) MOKE images of Pt/CoFeB/MgO after applying $I_p$ = 20 mA (top) or -20 mA (bottom). (h) MOKE images of Pt/Ni0.5 nm/CoFeB/MgO after $I_p$ = 25 mA (top) and -25 mA (bottom). (i) $t_{Ni}$ dependences of measured $H_{FL} + h_o$ and $I_c$ (-M to +M switching). The solid line is for eyes.

Current-induced magnetization switching was performed by first applying a current pulse ($I_P$) with the pulse length, $\tau_p$ = 1 ms, and then detecting the switched magnetization state through $\Delta R_{xx}$. Figures 3(c) and 3(d) show recorded $\Delta R_{xx}$ versus $I_p$ for Pt/CoFeB/MgO and Pt/Ni0.5nm/CoFeB/MgO, respectively, where typical current-induced magnetization switching behaviors are clearly presented. Considering opposite $\Delta R_{xx}$ signs, Figs. 3(c) and 3(d) indicate opposite switching directions, where +$I_P$ switches Pt/CoFeB/MgO to +M states but switches Pt/Ni0.5nm/CoFeB/MgO to -M states. Since $\tau_{FL}$ signs are opposite while $\tau_{DL}$ signs keep the same in the two samples, the reverse switching behaviors provide clearly evidences that the current-driven in-plane switching is induced by $\tau_{FL}$, rather than $\tau_{DL}$. The switching directions are also consistent with the measured $H_{FL} + h_O$. According to previous switching mechanisms, the switching direction was determined by $\tau_{DL}$ [1,3,4], which does not relate to ferromagnets and obviously conflicts present experimental observations. The switching directions were also examined directly by MOKE. Figure 3(e) shows the optical images of MOKE samples (width: 10 µm). The magnetization was first initialized to the -M state, and then, $I_p$ was applied. As expected, for Pt/CoFeB/MgO, +$I_p$ can switch magnetization from –M to +M states and -$I_p$ does not change the –M states (Fig. 3(g)); for Pt/Ni0.5nm/CoFeB/MgO, the –M states can only be switched by -$I_p$ (Fig. 3(h)). The domain formation and $\tau_{DL}$ influences on the $\tau_{FL}$-driven switching are discussed in Supplementary Materials.

Figure 3(i) shows $H_{FL} + h_O$ and critical switching current ($I_c$) versus $t_{Ni}$, where a clear correlation between magnetization switching and $\tau_{FL}$ is presented. First, $H_{FL}$ is very sensitive to $t_{Ni}$ and drops to 0.2 Oe with two monolayer Ni (approximate 0.4 nm) if $h_O$ = -0.7 Oe. It reduces to a negative value with increasing $t_{Ni}$ and $H_{FL} + h_O$ approaches around -1.6 Oe when 1 nm $\leq t_{Ni} \leq$ 2 nm. With larger $t_{Ni}$, $|H_{FL} + h_O|$ starts to reduce (Supplementary Materials). Therefore, $H_{FL}$ can be modulated at least one order of magnitude by controlling ferromagnets. Second, the switching direction is always consistent with the direction of $H_{FL} + h_O$. Positive $H_{FL} + h_O$ always switches magnetization to +M states, while negative $H_{FL} + h_O$ always switches to -M states. Third, $I_c$ is proportional to the efficiency of $\tau_{FL}$. When $H_{FL} + h_O$ approaches zero, no magnetization switching was observed until the sample was burnt (marked in Fig. 3(i)), even though considerable $\tau_{DL}$ can still be detected in this region (Fig. 2(e)). These results clearly demonstrate that the current-induced switching arises from $\tau_{FL}$ generated by ferromagnets. The $\tau_{FL}$-driven in-plane switching has also been verified in the W/Ni/CoFeB structures with an opposite SHE sign (Supplementary Materials). This also explain the inconsistent SOT efficiencies estimated from in-plane and perpendicular magnetization switching [4] because $\tau_{DL}$ dominates the latter [1,3,5,22].

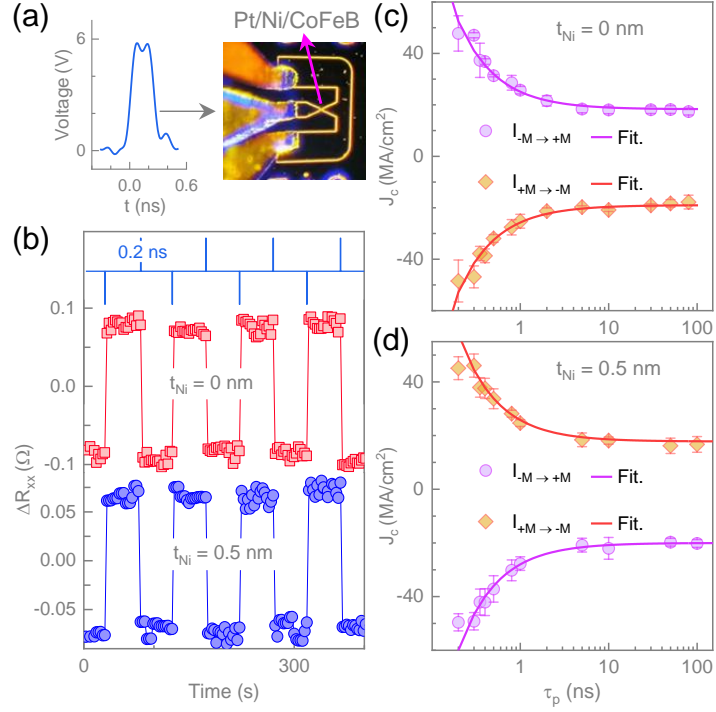

FIG. 4. (a) Optical images of CPW devices with RF probes for sub-nanosecond switching and the applied 0.2 ns pulses recorded by oscilloscopes. (b) Time dependence of applied 0.2 ns pulses (top) and corresponding measured $\Delta R_{xx}$ for Pt/CoFeB/MgO (middle) and Pt/Ni0.5nm/CoFeB/MgO (bottom). (c, d) The extracted $J_c$ versus $\tau_p$ for Pt/CoFeB/MgO (c) and Pt/Ni0.5nm/CoFeB/MgO (d). The solid lines are fitting results.

To demonstrate ultrafast $\tau_{FL}$-driven collinear switching without incubation time, the voltage pulses (0.2 ns to 100 ns) were applied to the samples through a coplanar waveguide (CPW) to reduce distortion. Figure 4(a) shows the applied 0.2 ns voltage pulses and optical images of CPW samples (Pt/Ni/CoFeB: 2.5 × 5 μm$^2$). Figure 4(b) shows the time dependence of magnetization switching induced by 0.2 ns write pulses. For both Pt/CoFeB/MgO and Pt/Ni0.5nm/CoFeB/MgO, the magnetization can be successfully switched by the 0.2 ns write pulses and the switching directions are consistent with Fig. 3. Figures 4(c) and (d) present the extracted critical switching current density ($J_c$) as a function of $\tau_p$, which can be well fitted by using $J_c = J_{c0} + q/\tau_p$. Here, $J_{c0}$ is the intrinsic $J_c$ and q describes the angular momentum transfer efficiency from charge to spin systems [43]. For Pt/CoFeB/MgO, fitting results give $J_{c0}$ = 18.36 MA/cm$^2$ (-M to +M) and -19.01 MA/cm$^2$ (+M to -M) with q = 8.54 × 10$^{-13}$ C. Remarkably, the switching time can be as short as that in $\tau_{DL}$-driven perpendicular switching without incubation time and $J_{c0}$ is about one order lower [43]. This experimentally confirms the simulated $\tau_{FL}$-driven switching process without incubation periods (Fig. 1(c)). The high $\tau_{FL}$ efficiency is also reflected in the charge-to-spin efficiency, where the extracted q value is about four times larger than that estimated from $\tau_{DL}$-driven switching [43] and also provides clear evidences to identify $\tau_{FL}$ as the driving force. For Pt/Ni0.5nm/CoFeB/MgO, $J_{c0}$ = -20.09 MA/cm² and 17.81 MA/cm² are determined with an even larger q = $1.10 \times 10^{-12}$ C, implying that the inserted Ni does change charge-to-spin efficiency for generating $\tau_{FL}$ as evidenced by direct $\tau_{FL}$ characterization (Fig. 2). These ultrafast switching measurements experimentally confirm that the write time of in-plane magnetized memory devices can be down to 0.2 ns, approaching the access time of commercial last-level cache memory, and more importantly, there is no the field-free switching issue encountered in perpendicular SOT memory [44,45]. Moreover, the energy efficiency and simple memory architecture also advance the $h_O$-driven toggle memory [46] (Supplementary Materials).

In summary, we have reported the observation of a ferromagnet-originated $\tau_{FL}$ that dominates the collinear in-plane switching. We show that both the strength and direction of $\tau_{FL}$ can be modulated independently with $\tau_{DL}$. The switching directions and ultrafast switching process down to 0.2 ns with high charge-to-spin conversion efficiency further confirm the $\tau_{FL}$-driven switching without incubation time. Compared to $\tau_{DL}$-driven perpendicular switching, it does not require an assisted magnetic field and thus may provide a practical solution for magnetic memory in competitive high-speed cache memory applications.

Supplemental Materials for

# Sub-nanosecond in-plane magnetization switching induced by field-like spin-orbit torques from ferromagnets

## 1: Sample fabrication and electrical measurements

The structures of Pt 2.5 nm/Ni $t_{Ni}$/CoFeB 2.5 nm/MgO 2 nm with 0 nm ≤ $t_{Ni}$ ≤ 2 nm were deposited on Si/SiO$_2$ (300 nm) substrates by magnetron sputtering with a base vacuum better than 2×10$^{-8}$ Torr. For $t_{Ni}$ = 0 nm, the structure of Pt 2.5 nm/ CoFeB 2.5 nm/MgO 2 nm was deposited directly without any Ni contamination. The other samples were deposited by using a wedged Ni layer. The samples were then patterned into a Hall bar structure with the width of 2.5 - 20 µm and the length of 50 µm. For ST-FMR measurements, the samples were patterned into a 20 µm microstrip first, and then a coplanar waveguide consisting of Cr 10 nm/Au 100 nm was deposited. The central part of Pt/Ni/CoFeB/MgO for microwave transmission is 20 µm × 30 µm with the length along the microwave current.

For USMR measurements, R$_{xx}$ was measured sequentially by using a positive and a negative dc current. $\Delta R_{xx} = R_{xx}(I) - R_{xx}(-I)$ was then calculated at each measurement step. Similarly, $\Delta R_{xy} = R_{xy}(I) - R_{xy}(-I)$ was measured to estimate field-like effective field and the Oersted field. To measure current-induced magnetization switching, a current pulse with 0.2 ns ≤ $\tau_p$ ≤ 1 ms was applied first. After waiting for 5 second, $\Delta R_{xx}$ was measured. During ST-FMR measuring, a microwave generator with the frequency capability of 2 GHz to 20 GHz was used to generate microwave current with the power of 20 dBm. The mix voltage due to FMR was separated through a bias tee and then detected by a voltmeter. The angle between the applied external magnetic field (H$_{ext}$) and microwave current was 45°.

## 2: Simulation of field-like torque switching

The simulation was performed via the stochastic Landau-Lifshitz-Gilbert (LLG) Equation:

$$\frac{d\mathbf{m}}{dt} = -\gamma\mu_0 \mathbf{m} \times (\mathbf{H}_{\text{eff}} + \mathbf{H}_{\text{r}}) + \alpha \mathbf{m} \times \frac{d\mathbf{m}}{dt} + \gamma\tau_{DL}\mathbf{m} \times (\mathbf{m} \times \hat{\mathbf{y}}) + \gamma\tau_{FL}\mathbf{m} \times \hat{\mathbf{y}} \qquad (1)$$

where $\alpha$ and $\gamma$ are the damping parameter and electron gyromagnetic ratio, respectively. $\mathbf{H}_{\text{eff}}$ represents the contributions of the applied field ($H_{app}$), exchange field, and magnetic anisotropic field ($H_k$). $\mathbf{H}_{\text{r}}$ represents the Gaussian fluctuating thermal field with magnitude $\sqrt{\frac{2\alpha k_B T}{\mu_0 M_s^2 V}}$, where $V$ is the volume of the ferromagnetic layer. We consider a ferromagnetic thin film with the easy axis along the y direction to perform the micromagnetic simulations. The x, y, and z directions, $\tau_{FL}$, and $\tau_{DL}$ are schematically shown in Fig. 1(a).

Micromagnetic simulations were performed by numerically solving the LLG equation at 300 K. Because of the thermal field, no initial angle is required, thus each run begins with the magnetization completely saturated along $+\hat{\mathbf{y}}$. The constant parameters in each set of runs are $T = 300$ K, $H_K = 4.0 \times 10^3$ A m$^{-1}$, $M_s = 1.0 \times 10^6$ A m$^{-1}$, A$_{ex}$ = 1.3 × 10$^{-11}$ J m$^{-1}$, and $\alpha = 0.02$. The sample dimension is 120 nm ×120 nm

×1.5 nm, and the unit cell size is 10 nm ×10 nm ×1.5 nm. In Figs. 1(b) and 1(c), $\tau_{FL} = 4.4 \times 10^3$ A m$^{-1}$ and $\tau_{DL} = 0$ A m$^{-1}$ were used for simulating field-like torque switching, and $\tau_{FL} = 0$ A m$^{-1}$ and $\tau_{DL} = 4.4 \times 10^3$ A m$^{-1}$ were used for damping-like torque switching.

### 3: Unidirectional magnetoresistance of CoFeB

As shown in Fig. S1, CoFeB shows strong USMR signal compared to cobalt that was used in other works [1]. Please note that, for Co samples, the measuring current (I) is ±1.5 mA (for other samples, ±1 mA) and USMR signal is enlarged by 15 times. In the large field range, the signs of $\Delta R_{xx}$ for all samples are the same, consistent with our understanding that USMR originates from spin accumulation at the interface. The spin accumulation arises from the spin Hall effects (SHE) in the Pt layer and the spin polarization direction does not depend on the ferromagnetic layer. Two peaks around zero field are opposite for CoFeB and Ni samples, which was attributed to the spin-magnon scattering [1]. However, the amplitude of USMR, $\Delta R = [\Delta R_{xx}(+H_y) - \Delta R_{xx}(+H_y)]/2$ as defined in Fig. S2(a), can be perfectly fitted by using a linear relationship at both high fields ($H_y = \pm 1000$ Oe) and low fields ($H_y = \pm 22$ Oe, a slight larger than coercivity, ~20 Oe), which does not show a $I^3$ dependence [1] (Fig. S2(b)). This means that the change of magnon population is not from pure thermal effects or the USMR mechanism presented in Ref. 1 is not valid for CoFeB and Ni. Interestingly, the USMR curve also depends on the annealing effect due to Joule heating after applying a large current pulse, as shown in Fig. S3. The annealing effect does not influence the conclusions of this work except increasing the magnitude of USMR around $H_y = 0$ Oe compered to as-deposited samples. The origin of USMR in these materials needs to be further confirmed in the future.

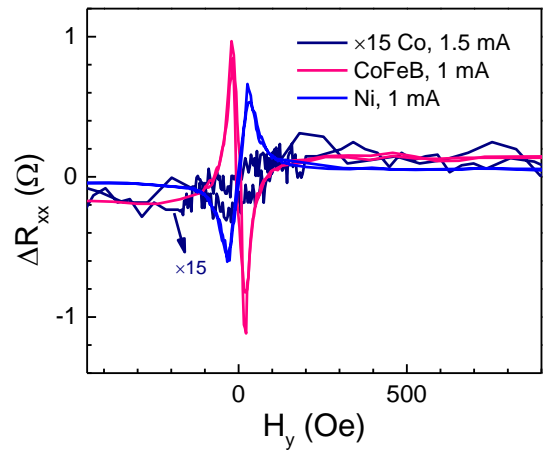

FIG. S1. USMR results for Pt 2.5 nm/Co 2 nm/MgO, Pt 2.5 nm/CoFeB 2.5 nm/MgO, and Pt 2.5 nm/Ni 5 nm/MgO. The measuring current for the Co sample is ±1.5 mA, and for CoFeB and Ni samples are ±1 mA. Please note that signal for the Co sample has been amplified by 15 times.

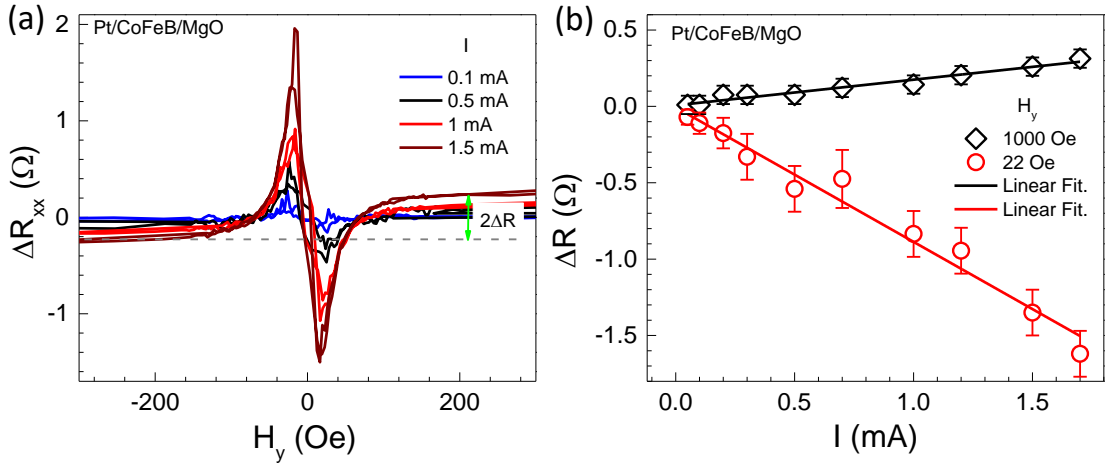

FIG. S2. (a) USMR results for Pt 2.5 nm/CoFeB 2.5 nm/MgO measured under different currents. The amplitude of USMR, $\Delta R = [\Delta R_{xx}(+H_y) - \Delta R_{xx}(+H_y)]/2$, is defined as indicated in the figure. (b) $\Delta R$ as a function of measuring current at $H_y = \pm 1000$ Oe and $\pm 22$ Oe. The solid lines are linear fitting results.

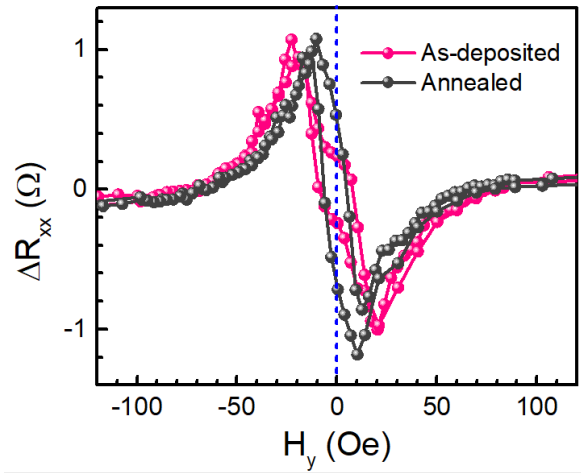

FIG. S3. Typical USMR results for Pt 2.5 nm/CoFeB 2.5 nm/MgO samples in as-deposited states and after annealing by applying +5 mA current pulses.

### 4: Current and Ni thickness dependences of field-like torque

Figure S4(a) shows $\Delta R_{xy}$ as function of $H_x$ under different applied currents without compensated field ($h_y = 0$) for Pt 2.5 nm/CoFeB 2.5 nm/MgO samples, which reflects the current-induced $\tau_{FL}$. The two peaks around zero field can be eliminated by applying a negative compensated field. Figure S4(b) shows the measured current-induced effective field, $H_{FL} + h_O$, as a function of applied current. As expected, the current-induced effective field is proportional to the applied current and can be fitted by using a linear relationship. Fig. S5 shows the measured $H_{FL} + h_O$ as a function of $t_{Ni}$ measured at I = ± 1 mA. When $t_{Ni} \geq$ 2.0 nm, $|H_{FL} + h_O|$ decreases with increasing $t_{Ni}$ but not approaches a constant value because the current flowing through the Pt layer that contributes $h_O$ also change with $t_{Ni}$.

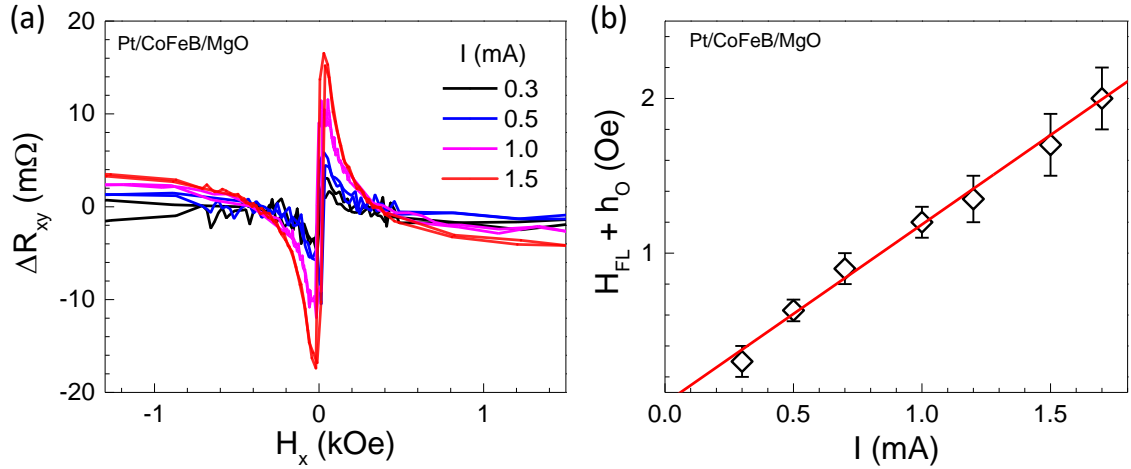

FIG. S4. (a) $\Delta R_{xy}$ as a function of $H_x$ for Pt 2.5 nm/CoFeB 2.5 nm/MgO measured under different currents. (b) Current-induced effective field determined by a compensated magnetic field as a function of applied current, which can be linearly fitted as shown by the solid line.

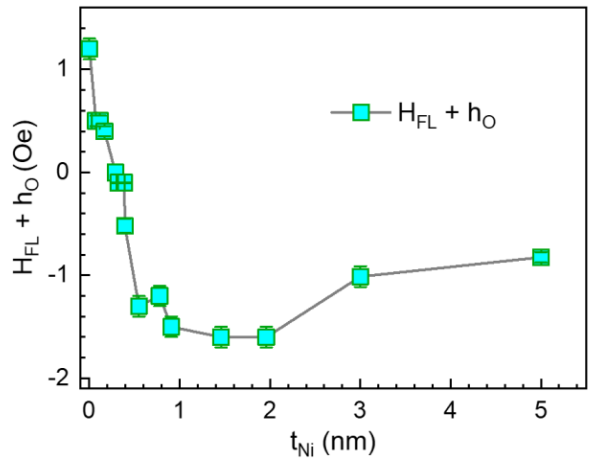

FIG. S5. The measured $H_{FL} + h_O$ as a function of $t_{Ni}$ in Pt 2.5 nm/Ni ($t_{Ni}$)/CoFeB 2.5 nm/MgO structures.

**5: Calibration of Oersted field in Pt/CoFeB structures**

For calibration of the Oersted field ($h_O$), an inverse sample with the structure of Si/SiO$_2$ 300 nm/CoFeB 2.5 nm/MgO 2 nm/Pt 2.5 nm/MgO 2 nm was deposited and then patterned into a Hall bar structure with the same dimension as Sub/Pt 2.5 nm/CoFeB 2.5 nm/MgO 2 nm. The inserted 2 nm MgO between the CoFeB and Pt layers is used to forbid any spin current injection to the CoFeB layer while keeping all the conducting layers exactly the same as Sub/Pt 2.5 nm/CoFeB 2.5 nm/MgO 2nm samples. Therefore, it is reasonable to assume that $h_O$ generated from the Pt layer has the same value but with opposite directions for the two samples. We still use the field-compensated method to detect $h_O$. Figure S6(a) shows $\Delta R_{xy}$ as function of $H_x$ under different compensated field ($h_y$) for Si/SiO$_2$ 300nm/CoFeB 2.5 nm/MgO 2 nm/Pt 2.5 nm/MgO 2 nm. Noted that the applied current is ± 1.5 mA. Compared to the Sub/Pt 2.5 nm/CoFeB 2.5 nm/MgO 2 nm samples shown in Fig. 2(a) in main text, the two peaks around zero field have the same direction when $h_y = 0$ and can be modulated and even reversed by a negative $h_y$. When $h_y = -1.0$ Oe, the two peaks disappear, indicating $h_O = 1.0$ Oe. In the high field range, $\Delta R_{xy} \propto \boldsymbol{M} \times \Delta T$ (arising from the anomalous Nernst effect) becomes reverse compared to Fig. 2(a). This is because the temperature gradient $\Delta T$ in this control sample is reversed compared to Sub/Pt 2.5 nm/CoFeB 2.5 nm/MgO 2 nm samples by considering that the Joule heating is mainly from the Pt layer. Figure S6(b) shows the measured $h_O$ as a function applied current, which can also be fitted perfectly by a linear function. From this fitting, we determined that $h_O \approx 0.7$ Oe when I = 1 mA in this structure which is used for calculating $H_{FL}$ in Sub/Pt 2.5 nm/CoFeB 2.5 nm/MgO 2 nm samples ($h_O \approx -0.7$ Oe in Sub/Pt 2.5 nm/CoFeB 2.5 nm/MgO 2 nm samples due to inverse structures). Please note that the measured Oersted field is much smaller than that calculated by $h_O = I/(2w)$ where $w$ is the width of Hall bar. The possible reason can be that the width of Hall bar (along the current direction) in our samples is 2.5 µm and the width of voltage bar (orthogonal to the current direction) is 2 µm, in which the applied current is not exactly confined in the width of 2.5 µm as shown in inset of Fig. S6(b).

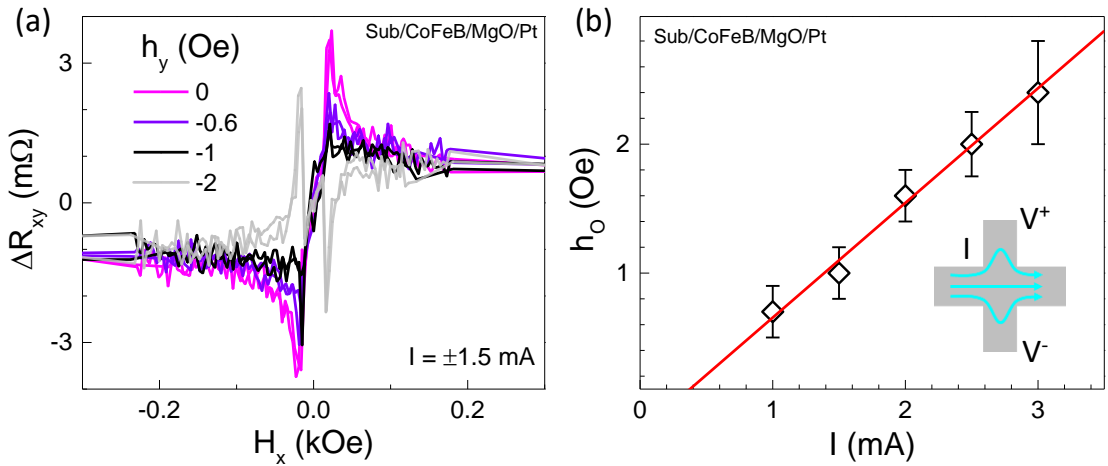

FIG. S6. (a) $\Delta R_{xy}$ as a function of $H_x$ for Sub/CoFeB 2.5 nm/MgO 2 nm/Pt 2.5 nm/MgO with different compensated field $h_y$. The measuring current is ±1.5 mA. (b) the measured Oersted field as a function of applied current, which can be fitted linearly. Inset shows the schematic of current flowing through the Hall bar structure. The arrows indicate current direction.

## 6: Field-like torque driven in-plane switching in W/Ni/CoFeB structures

We have also verified that the reversal of $\tau_{FL}$ and in-plane switching directions in W 6 nm/Ni $t_{Ni}$/CoFeB 2.5 nm/MgO 2nm structures with a negative spin Hall angle (SHA). Similar to the Pt-based structures with a positive SHA, we first observed the sign change of $\tau_{FL}$ with the Ni insertion by using the field-compensated measurements, as shown in Fig. S7(a) and S7(b). Without Ni insertion, the measured $H_{FL}$ + $h_O$ is negative, while with 1.2 nm Ni insertion, $H_{FL}$ + $h_O$ becomes positive. Then, we demonstrated the in-plane magnetization switching can be detected by recording USMR signals in W/Ni/CoFeB systems, which shows that $\Delta R_{xx}$ corresponding to the same magnetized direction also changes sign with the Ni insertion, as shown in Fig. S7(c) and S7(d). Finally, as shown in Fig. S7(e) and S7(f), we observed the similar current-driven in-plane magnetization switching like Pt/Ni/CoFeB structures, where the same switching directions of $\Delta R_{xx}$ curves indicate the magnetization switching directions are opposite with/without the Ni insertion. These results demonstrate that $\tau_{FL}$ also dominates the in-plane magnetization switching in W/Ni/CoFeB systems with a negative SHA.

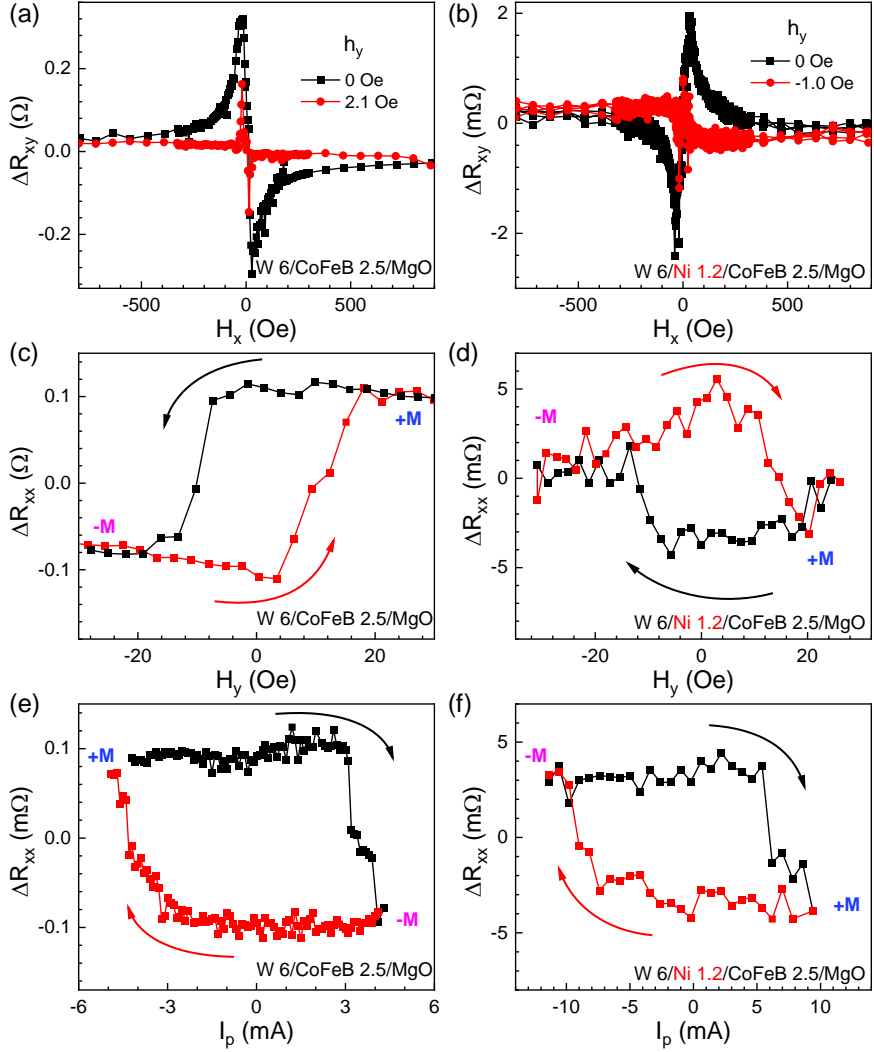

FIG. S7. (a, b) $\Delta R_{xy}$ as a function of $H_x$ under $h_y$ modulations for (a) W 6 nm/CoFeB 2.5 nm/MgO 2 nm ($H_{FL} + h_O = -2.1$ Oe) and (b) W 6 nm/Ni 1.2 nm/CoFeB 2.5 nm/MgO 2 nm ($H_{FL} + h_O = +1.0$ Oe). (c, d) Magnetic field-driven in-plane switching detected by USMR signals for (c) W 6 nm/CoFeB 2.5 nm/MgO 2 nm and (d) W 6 nm/Ni 1.2 nm/CoFeB 2.5 nm/MgO 2 nm. (e, f) Current-driven in-plane switching in (e) W 6 nm/CoFeB 2.5 nm/MgO 2 nm and (f) W 6 nm/Ni 1.2 nm/CoFeB 2.5 nm/MgO 2 nm samples. The dimension of Hall bars: 20 μm × 2 μm; Applied current: (a): ± 1.5 mA; (b): ± 3.0 mA.

## 7: No current-driven in-plane switching in Cu/CoFeB structures

In general, $h_O$ can also induce in-plane magnetization switching without $H_{FL}$ as adopted in toggle MRAM. However, the toggle MRAM requires the thickness of write bit lines generating $h_O$ is larger than 300 nm [2]. To demonstrate if $h_O$ alone can switch the in-plane magnetization in ultrathin metal/ferromagnet bilayers, we fabricated Cu 5 nm/CoFeB 2.5 nm/MgO 2 nm samples (Hall bar patterns, 3 × 30 μm$^2$), where $H_{FL}$ is not expected. Fig. S8(b) shows the measured $H_{FL} + h_O$ ($H_{FL}$ should be 0) as a function of current using the same measurements as Pt/CoFeB samples (Fig. 2a-2c). First, we confirm that $H_{FL} + h_O$ is opposite in Cu/CoFeB and Pt/CoFeB samples if we assume that only $h_O$ exists in the Cu/CoFeB samples. Second, the current-generated $H_{FL} + h_O$ efficiency is much lower in the Cu/CoFeB samples as shown in Fig. S9 (slope for the Hall bar width of 2.5 μm: 1.16 Oe/mA for Pt/CoFeB is about two times larger than -0.64 Oe/mA for Cu/CoFeB). Third, similar to Fig. 3a, we can also observe field-driven in-plane magnetization switching in the Cu/CoFeB samples, as shown in Fig. S8(c) (USMR exists even in a single CoFeB layer [3], and thus, the in-plane magnetization states can still be detected in Cu/CoFeB samples). However, no current-driven in-plane magnetization switching was observed until the devices were burnt by Joule heating as shown in Fig. S8(d). According to Fig. S8(c), the y-directional switching field of Cu/CoFeB devices is about ±10 Oe; To generate $h_O = \pm 10$ Oe, the estimated applied current is about 18.9 mA according to Fig. S8(b). However, Fig. S8(d) shows that the maximum tolerant current where the devices are burnt by Joule heating is about 6.5 mA, much smaller than the required current (18.9 mA) for generating $h_O = \pm 10$ Oe. Therefore, it is reasonable that $h_O$ alone cannot switch the in-plane magnetization in the ultrathin (several nm) metal/ferromagnet bilayers. Apparently, $h_O$ can switch in-plane magnetization without burning in toggle MRAM because the thicker metal layers can tolerate large current to generate $h_O$ for switching.

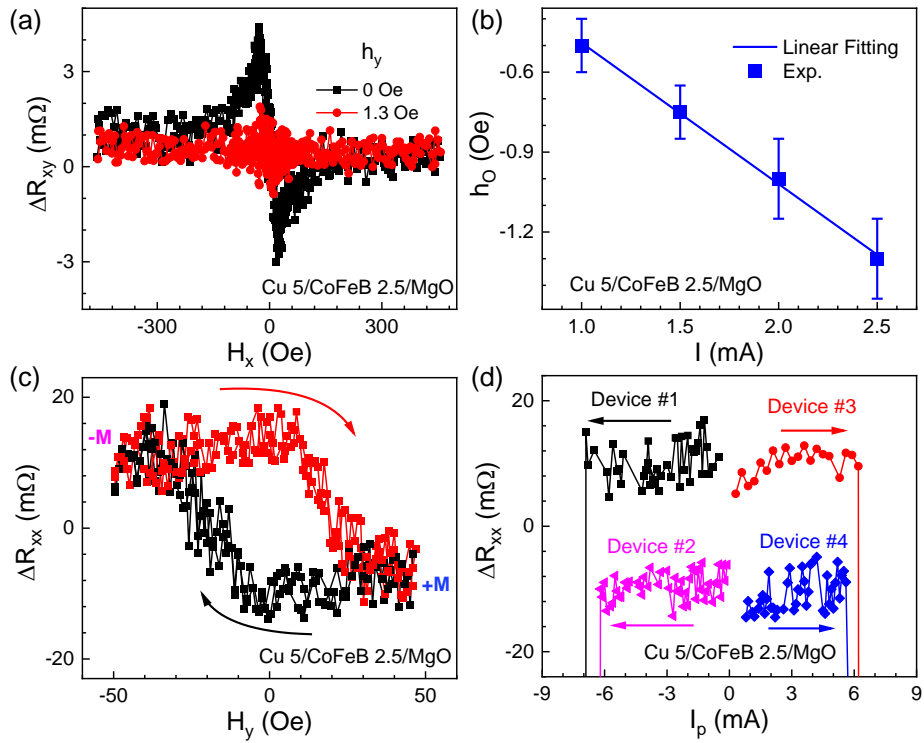

FIG. S8. (a) $\Delta R_{xy}$ as a function of $H_x$ under $h_y$ modulations for Cu 5 nm/CoFeB 2.5 nm/MgO (Hall bar dimension: 30 μm × 3 μm; current: ±2.5 mA). (b) The measured $h_O$ shows a linear dependence on the applied current. (c) Magnetic field-driven in-plane switching detected by USMR signals for Cu 5 nm/CoFeB 2.5 nm/MgO. (d) No current-driven in-plane switching until the devices were burnt by applied current.

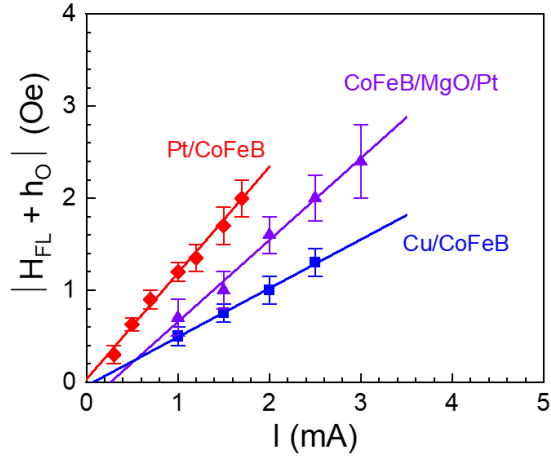

FIG. S9. The measured ∣$H_{FL}+h_O$∣ as a function of applied current for Pt 2.5 nm/CoFeB 2.5 nm/MgO (Hall bar width: 2.5 μm), CoFeB 2.5 nm/MgO 2 nm/Pt 2.5 nm (Hall bar width: 2.5 μm), and Cu 5 nm/CoFeB 2.5 nm/MgO (Hall bar width: 3 μm).

## 8: Damping-like torque influences on the field-like torque driven switching

Since Fig. 2(d) and 2(f) show that $\tau_{FL}$ changes sign while $\tau_{DL}$ does not, we simulated the current-driven in-plane switching when $\tau_{FL}$ and $\tau_{DL}$ have the same or opposite signs. As shown in Fig. S10, $\tau_{DL}$ can significantly influence the $\tau_{FL}$-driven switching: when $\tau_{FL}$ and $\tau_{DL}$ have the same sign, $\tau_{DL}$ assists the $\tau_{FL}$ switching and the switching time becomes short; when they have opposite signs, $\tau_{DL}$ hampers the $\tau_{FL}$ switching and the switching time becomes long. Therefore, in practical applications, by modulating the SOT-material/ferromagnet interfaces, $\tau_{FL}$ and $\tau_{DL}$ can be designed to have the same sign and work collectively to accelerate the current-driven in-plane switching processes.

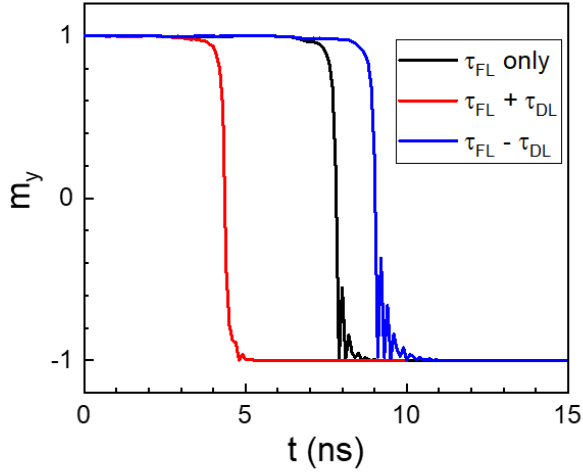

FIG. S10. The simulated $\tau_{FL}$-driven switching with/without $\tau_{DL}$. $\tau_{FL} = 1.4 \times 10^3$ A m$^{-1}$ and $\tau_{DL} = \pm 1.4 \times 10^3$ A m$^{-1}$.

## 9: Domain nucleation and expansion in the field-like torque driven switching

The domain nucleation and following domain expansion are the general switching behaviors for the large devices where the domains can be nucleated. Generally, the first switching region (domain nucleation) is induced by the same mechanisms as small devices showing coherent switching, and the switching behaviors like switching directions and mechanisms are not expected to change with the device size. However, the critical switching field/current and switching time can be significantly influenced by domain nucleation and domain wall motion. To investigate the domain nucleation and expansion in the $\tau_{FL}$-driven switching, we simulated the switching process by using 120 nm (Fig. S11) and 1.5 μm (Fig. S12-14) devices in three cases: (I) only $\tau_{FL}$ (Fig. S11 and S12); (II) $\tau_{FL}$ and $\tau_{DL}$ with the same sign (Fig. S13); (III) $\tau_{FL}$ and $\tau_{DL}$ with opposite signs (Fig. S14). As expected, for the 120 nm devices, there is no domain formation during the switching processes, while for the 1.5 μm devices, the switching is ignited by domain nucleation and then finished through domain expansion. In all the three cases, the switching directions are not influenced by the dimension of devices but the switching process is much faster for the 1.5 μm devices. $\tau_{DL}$ assisting/hampering $\tau_{FL}$ switching with the same/opposite signs is also reflected in the 1.5 μm devices with domain nucleation and expansion, as shown in Fig. S12-14. One can see that the switching processes of current-driven in-plane switching in large devices are different from that of current-driven perpendicular switching, where the domain wall motion may determine the switching direction [4].

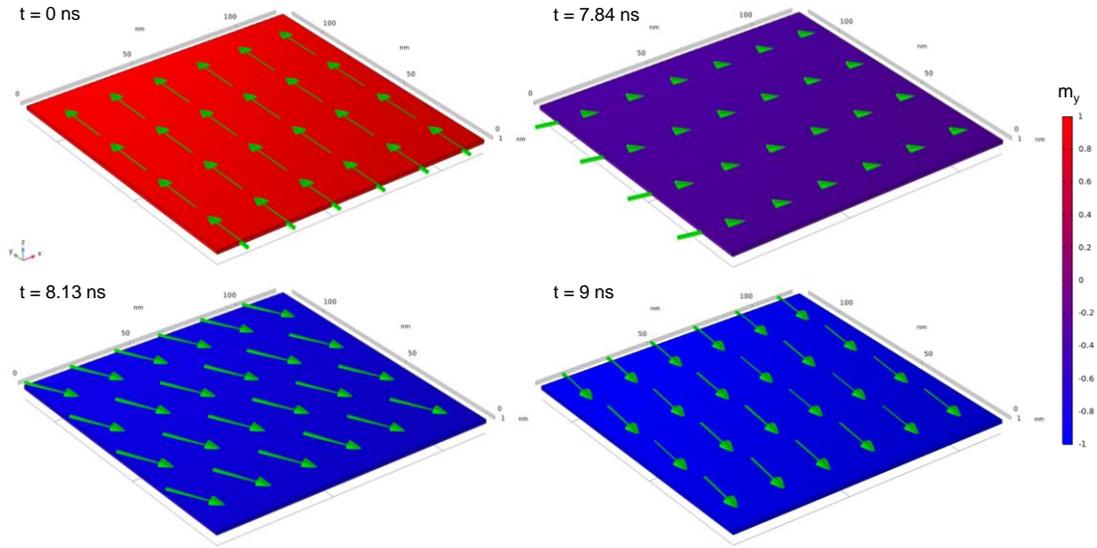

FIG. S11. The simulated $\tau_{FL}$-driven switching processes in 120 nm × 120 nm devices. $\tau_{FL} = 1.4 \times 10^3$ A m$^{-1}$ and $\tau_{DL}$ = 0 A m$^{-1}$.

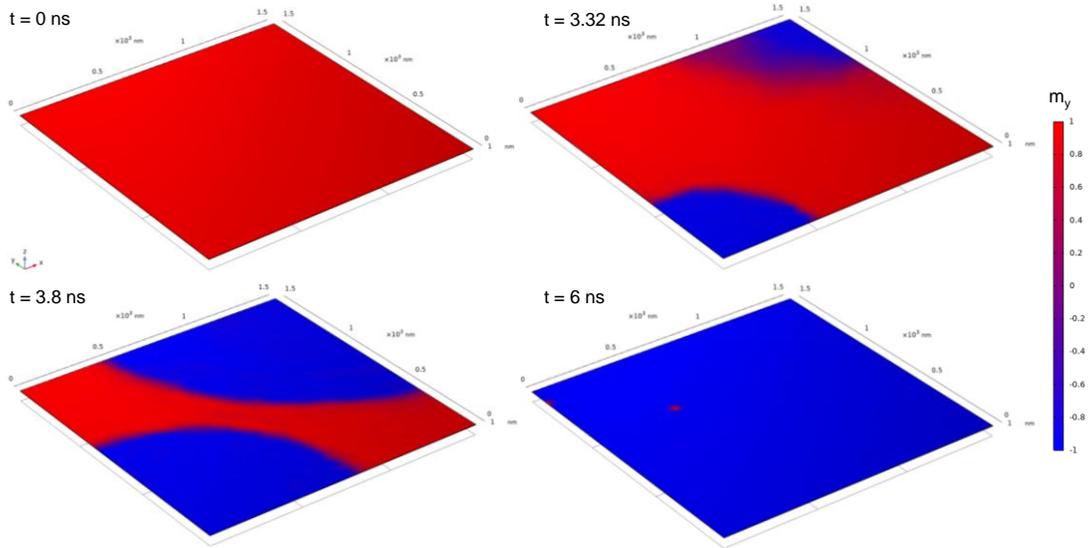

FIG. S12. The simulated $\tau_{FL}$-driven switching processes in 1.5 μm × 1.5 μm devices. $\tau_{FL} = 1.4 \times 10^3$ A m$^{-1}$ and $\tau_{DL}$ = 0 A m$^{-1}$.

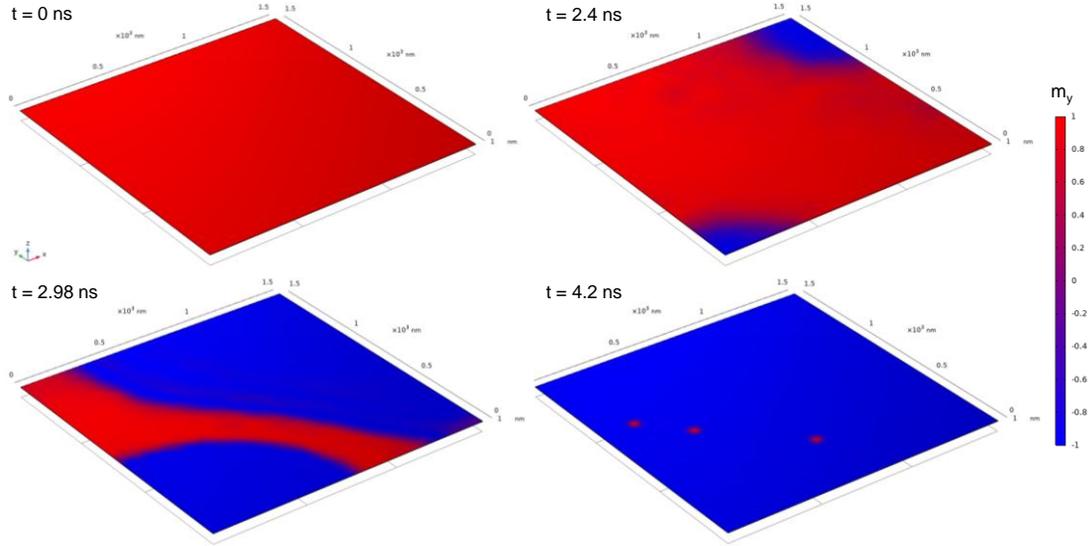

FIG. S13. The simulated current-driven in-plane switching processes in 1.5 μm × 1.5 μm devices when $\tau_{FL} = 1.4 \times 10^3$ A m$^{-1}$ and $\tau_{DL} = 1.4 \times 10^3$ A m$^{-1}$.

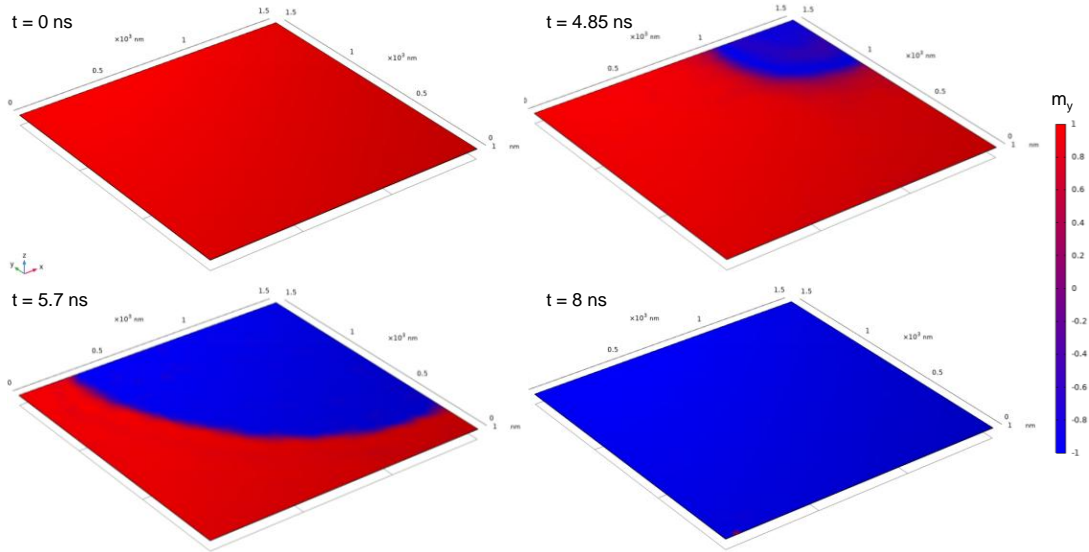

FIG. S14. The simulated current-driven in-plane switching processes in 1.5 μm × 1.5 μm devices when $\tau_{FL} = 1.4 \times 10^3$ A m$^{-1}$ and $\tau_{DL} = -1.4 \times 10^3$ A m$^{-1}$.